# Modeling of graphene-based NEMS

Irina V. Lebedeva[a,b,c,1], Andrey A. Knizhnik[b,c], Andrey M. Popov[d], Yurii E. Lozovik[a,d] and Boris V. Potapkin[b,c]

[a]Moscow Institute of Physics and Technology, Institutskii pereulok 9, Dolgoprudny, Moscow Region 141701, Russia

[b]National Research Centre "Kurchatov Institute", Kurchatov Square 1, Moscow 123182, Russia

[c]Kintech Lab Ltd., Kurchatov Square 1, Moscow 123182, Russia

[d]Institute of Spectroscopy of Russian Academy of Sciences, Fizicheskaya Street 5, Troitsk, Moscow Region 142190, Russia

**Abstract**

The possibility of designing nanoelectromechanical systems (NEMS) based on relative motion or vibrations of graphene layers is analyzed. *Ab initio* and empirical calculations of the potential relief of interlayer interaction energy in bilayer graphene are performed. A new potential based on the density functional theory calculations with the dispersion correction is developed to reliably reproduce the potential relief of interlayer interaction energy in bilayer graphene. Telescopic oscillations and small relative vibrations of graphene layers are investigated using molecular dynamics simulations. It is shown that these vibrations are characterized with small Q-factor values. The perspectives of nanoelectromechanical systems based on relative motion or vibrations of graphene layers are discussed.

---

[1] Corresponding author. Tel./Fax: +7 499 196 9992. *E-mail address*: lebedeva@kintechlab.com (I.V. Lebedeva).

## 1. Introduction

Due to the unique electronic and mechanical properties of carbon nanostructures, they have been considered as promising materials for a variety of applications. In addition to zero-dimensional and one-dimensional carbon nanostructures, fullerenes and carbon nanotubes, a novel two-dimensional carbon nanostructure, graphene, was discovered recently [1]. Most of the studies on graphene have been focused on its electronic properties. However, there were also a few works devoted to the mechanical properties of graphene [2, 3]. In particular, a self-retracting motion of graphite, i.e. retraction of graphite flakes back into graphite stacks on their extension, was observed experimentally [3]. This phenomenon is similar to the self-retracting motion of nanotube walls [4] arising from their van der Waals interaction. The ability of free relative sliding and rotation of carbon nanotube walls [4, 5] and their excellent "wearproof" characteristics [5] allowed using carbon nanotube walls as movable elements in nanoelectromechanical systems (NEMS). By analogy with the gigahertz oscillator based on carbon nanotubes [6, 7], a gigahertz oscillator based on the telescopic oscillation of graphene layers was suggested [3].

The possibilities to use carbon nanotubes in miniature devices have been thoroughly studied. A number of devices offering great promise for applications in NEMS and based on the use of the relative motion or vibrations of carbon nanotube walls have been proposed recently. These devices include rotational [8, 9] and plain [4] nanobearings, nanoactuators [10, 11], Brownian motors [12], nanobolt-nanonut pairs [13–15], gigahertz oscillators [6, 7] and ultra-high frequency nanoresonators [16]. Furthermore, nanomotors based on the relative rotation of carbon nanotube walls [17–20] and memory cells based on the telescopic extension of carbon nanotube walls [21, 22] were implemented.

Up to now only a few works [2, 3] have been devoted to investigation of graphene-based NEMS. Though carbon nanotubes and graphene are close in their honeycomb structure, graphene shows a number of differences which affect the possibility to use it in NEMS. While potential reliefs of the interwall interaction energy of carbon nanotubes and, consequently, the characteristics of the nanotube-based NEMS can be varied in a wide range depending on the chiralities and radii of the nanotube walls [16, 23], the graphene-based NEMS should not display such a variety of the properties. The magnitude of corrugation of the interaction energy against sliding for commensurate graphene layers [24, 25] greatly exceeds the maximal values for carbon nanotubes [16, 26, 27] reached for commensurate non-chiral nanotube walls. This should provide different characteristics of the graphene-based NEMS compared to the nanotube-based NEMS. In the present paper, we analyze the possibilities to design the gigahertz oscillator [3] based on the telescopic oscillation of graphene layers, a nanorelay based on the telescopic motion of graphene layers and a nanoresonator based on the small relative vibrations of graphene layers, which can be proposed by analogy with the carbon nanotube-based devices [6, 7, 16, 21, 22].

The characteristics of the graphene-based NEMS are determined by the potential relief of interaction energy of graphene layers. We perform density functional theory calculations with the dispersion correction (DFT-D) to obtain the potential relief of interlayer interaction energy in bilayer graphene. Based on these calculations, we develop a classical potential, which reliably reproduces the magnitude of corrugation of the potential relief, barrier to relative motion of graphene layers and frequency of small relative vibrations of the layers. The developed potential is applied for molecular dynamics (MD) simulations of the telescopic oscillation and small relative vibrations of graphene layers. We show both the telescopic oscillation and small relative vibrations of graphene layers are characterized by large damping. On the one hand, this implies that graphene is not a suitable material for the gigahertz oscillator and nanoresonator. On the other hand, the large damping allows elaborating the nanorelays and memory cells based on the telescopic motion of graphene layers which are fast-responding due to the absence of mechanical oscillations after switching.

## 2. Analysis of potential relief of graphene interlayer energy

To analyze the operation of NEMS based on the relative motion or vibrations of graphene layers, it is required to reliably reproduce the potential relief of interaction energy of graphene layers. We performed calculations of the interlayer interaction energy in bilayer graphene using empirical and *ab initio* methods. In the calculations of the potential energy reliefs, one of the graphene layers was rigidly shifted parallel to the other. Account of structure deformation induced by the interlayer interaction was shown to be inessential for the shape of the potential relief of interaction energy between graphene-like layers, such as the interwall interaction of carbon nanotubes [9, 24] and the intershell interaction of carbon nanoparticles [28, 29].

The recently developed DFT-D method [30, 31], which enables taking into account the van der Waals interaction in the DFT calculations, was used to obtain the potential relief of the interlayer interaction energy in bilayer graphene with high accuracy. The periodic boundary conditions were applied to a 4.26 Å x 2.46 Å x 20 Å model cell. The VASP code [32] with the generalized gradient approximation (GGA) density functional of Perdew, Burke, and Ernzerhof [33] corrected with the dispersion term (PBE-D) [34] was used. The basis set consisted of plane waves with the maximum kinetic energy of 500 eV. The interaction of valence electrons with atomic cores was described using the projector augmented-wave method (PAW) [35]. Integration over the Brillouin zone was performed using 24 x 36 x 1 k-point sampling.

The interaction energy of the graphene layers, equilibrium interlayer spacing and bulk modulus obtained using the DFT-D calculations are listed in Table 1. The calculated interlayer interaction energy in bilayer graphene as a function of the relative displacement of the graphene layers at the equilibrium interlayer spacing is shown in Fig. 1. The found minimum energy states correspond to the AB-stacking of the layers, while the maximum energy states correspond to the AA-stacking, in agreement with the experiment [36]. The SP-stacking corresponds to the energy barrier for transition of the layers between adjacent energy minima represented by the AB stacking. The relative energies of the AA, SP and AB stackings are given in Table 1. The frequency $f_0$ of small relative vibrations of the layers about the global energy minimum is also presented in Table 1.

Table 1. Characteristics of the potential relief of interlayer interaction energy in bilayer graphene: interlayer binding energy $E_{AB}$, interlayer spacing $c_0/2$, bulk modulus $B$, relative energy $E_{AA} - E_{AB}$ of the AA stacking, relative energy $E_{SP} - E_{AB}$ of the SP stacking, frequency $f_0$ of relative translational vibrations of the graphene layers.

|  | DFT-D | New potential | Experimental data |
|---|---|---|---|
| $E_{AB}$ (meV per atom) | −50.6 | −46.9 | −52±5[a], 43[b], 35[c] |
| $c_0/2$ (Å) | 3.25 | 3.37 | 3.328[d], 3.354[e] |
| $B$ (GPa) | 38.9 | 38.3 | 41[f] |
| $E_{AA} - E_{AB}$ (meV per atom) | 19.5 | 19.5 |  |
| $E_{SP} - E_{AB}$ (meV per atom) | 2.07 | 2.07 |  |
| $f_0$ (THz) | 1.04 | 1.06 | 0.95 (1.35[g]) |

[a]Ref. [37]; [b]Ref. [38]; [c]Ref. [39]; [d]Ref. [40]; [e]Ref. [41]; [f]Ref. [42]; [g] Estimated on the basis of the data for graphite from Ref. [43] (given in parentheses)

As mentioned in Ref. [34], the PBE-D functional closely reproduces the experimental data on the interlayer binding energy [37–39], interlayer spacing [40, 41] and bulk modulus [42] of graphite. Moreover, according to our calculations, it provides the frequency of the small relative vibrations of graphene layers in bilayer graphene close to the estimate of 0.95 THz derived from the experimental frequency of the TO' mode of graphite at Γ-point of about 1.35 THz [43] (see Table 1). The calculated magnitude of corrugation against sliding for commensurate graphene layers exceeds the maximal values for carbon nanotubes [16, 26, 27] reached for commensurate non-chiral nanotube walls by one-two orders of magnitude. This is due to the perfect matching between the graphene layers as opposed to the curved nanotube walls.

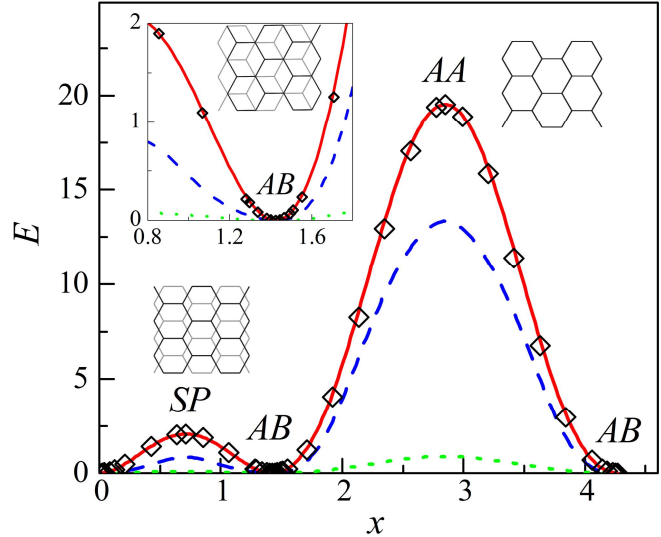

Fig. 1. Calculated interaction energy (in meV per atom) of graphene layers at the equilibrium interlayer spacing as a function of the relative displacement $x$ (in Å) of the layers along the armchair direction for different potentials: Lennard-Jones potential (green dotted line), Kolmogorov-Crespi potential (blue dashed line) and developed potential (red solid line). The data obtained from the DFT-D calculations are shown by diamonds. The energy is given relative to the global energy minimum.

Based on the data obtained by the DFT-D calculations, we have developed a classical potential in the form similar to the one suggested lately by Kolmogorov and Crespi [24, 25]. It was pointed out in papers [24, 25] that the $\pi$-overlap between graphene layers is anisotropic. So to fit both the experimental graphite compressibility and the corrugation against sliding, it is needed to distinguish the in-plane and out-of-plane directions. We assumed that the interaction of atoms of the layers at distance $r$, transverse separation $\rho$ and interlayer spacing $z$ ($r^2 = \rho^2 + z^2$) can be described as

$$U = A\left(\frac{z_0}{r}\right)^6 + B\exp(-\alpha(r-z_0)) +$$
$$C\left(1 + D_1\rho^2 + D_2\rho^4\right)\exp(-\lambda_1\rho^2)\exp(-\lambda_2(z^2 - z_0^2))$$
(1)

Using expression (1), the interlayer binding energy [37–39], interlayer spacing [40, 41] and bulk modulus [42] of graphite were closely fitted to the recent experimental values (see Table 1). The dependence of the potential energy on the relative displacement of the graphene layers was fitted to the results of the DFT-D calculations (see Fig. 1, Table 1). The parameters of the potential were found to be $A = -10.510$ meV, $B = 11.652$ meV, $\alpha = 4.16$ Å$^{-1}$, $C = 35.883$ meV, $D_1 = -0.86232$ Å$^{-2}$, $D_2 = 0.10049$ Å$^{-4}$, $\lambda_1 = 0.48703$ Å$^{-2}$, and $\lambda_2 = 0.46445$ Å$^{-2}$. The cutoff distance was taken equal to 25 Å. The root-mean square deviation of the potential energy relief calculated using the fitted potential (3) from the relief obtained using the DFT-D calculations is only 0.16 meV/atom. The frequency of small relative vibrations of the graphene layers exceeds the DFT-D value by about 2% and is rather close to the experimental estimate [43] (see Table 1).

We also repeated the calculations of the potential relief of interlayer interaction energy in bilayer graphene using the Kolmogorov-Crespi potential [25] (see Fig. 1). As opposed to the developed potential, the Kolmogorov-Crespi potential displays a significant deviation from the potential energy relief obtained using the present DFT-D calculations of about 2.6 meV/atom (see Fig. 1).

Furthermore, the Lennard-Jones potential ($U_{LJ} = 4\varepsilon\left((\sigma/r)^{12} - (\sigma/r)^6\right)$) was also considered for comparison. The parameters of the Lennard-Jones potential $\varepsilon = 2.757$ meV, $\sigma = 3.393$ Å were fitted by us to reproduce the interlayer binding energy, interlayer spacing and bulk modulus of graphite. The cut-off distance of the potential was equal to 25 Å. The Lennard-Jones potential provides satisfactory values of the interlayer binding energy, interlayer spacing and bulk modulus for graphite. However, the magnitude of corrugation of the potential energy relief is underestimated by an order of magnitude (see Fig. 1).

## 3. MD simulations of graphene-based NEMS

Similar to the ultra-high frequency nanoresonator based on the small relative vibrations of carbon nanotube walls [16], a nanoresonator based on the small relative vibrations of graphene layers can be considered. The frequency of such vibrations lies in the range of $f = 0.2 – 1.1$ THz for the potential reliefs of interlayer interaction energy calculated using the empirical and *ab initio* methods (see Table 1). This is of the order of the values predicted for double-walled carbon nanotubes [16]. However, to analyze possible applications of the nanoresonator the dynamic behavior of the system should be investigated.

In the MD simulations, we considered two systems consisting of two infinite graphene layers and of a graphene flake on an infinite graphene layer. To model the infinite layers the periodic boundary conditions were applied along mutually perpendicular armchair and zigzag directions. The size of the model cell was 5.1 nm x 5.2 nm, respectively. The size of the graphene flake was 2.0 nm along the armchair edge and 2.1 nm along the zigzag edge (178 carbon atoms). The covalent carbon-carbon interactions in the layers were described by the empirical Brenner potential [44], which was shown to correctly reproduce the vibrational spectra of carbon nanotubes [45] and graphene nanoribbons [46] and has been widely applied to study carbon systems [16, 23, 47, 48]. Microcanonical MD simulations of relative vibrations of the layers were performed at liquid helium temperature of 4.2 K. An in-house MD-kMC code [49] was implemented. The time step was 0.4 fs. To start the vibration one of the layers was shifted by 0.2 Å from the energy minimum in the armchair direction and released with the zero center-of-mass velocity. During the simulations, both layers were free.

The calculated relative displacement of the centers of mass of the layers as a function of time is shown in Fig. 2. To estimate the frequency and Q-factor of these vibrations, the Fourier transform of the relative displacement of the layers was obtained (see Fig. 3). The frequency $f$ of the vibrations was found as the center of the main peak and the Q-factor was estimated by the width $\Delta f$ of the peak as

$$Q = \frac{f}{2\pi \Delta f}. \tag{2}$$

The frequencies and Q-factors obtained by the MD simulations using different potentials and systems are listed in Table 2. As it is seen from Fig. 3 and Table 2, the Q-factor of the graphene-based nanoresonator is relatively small $Q < 200$ for all the potentials. The frequency of vibrations of the flake is smaller than the frequency of relative vibrations of the infinite layers (see Fig. 3 and Table 2) due to the lower ratio of the steepness of the interlayer interaction energy to the reduced mass. Because of the additional relaxation of the flake relative to the infinite layer constrained by the boundary conditions, this decrease of frequency is smaller for the developed potential and Kolmogorov-Crespi potential than for the Lennard-Jones potential. For the developed potential and the Kolmogorov-Crespi potential, there is also a dramatic decrease in the Q-factor for the graphene flake on the infinite graphene layer compared to the two infinite layers (see Fig. 3 and Table 2), which should be attributed the enhanced dissipation at the flake edges.

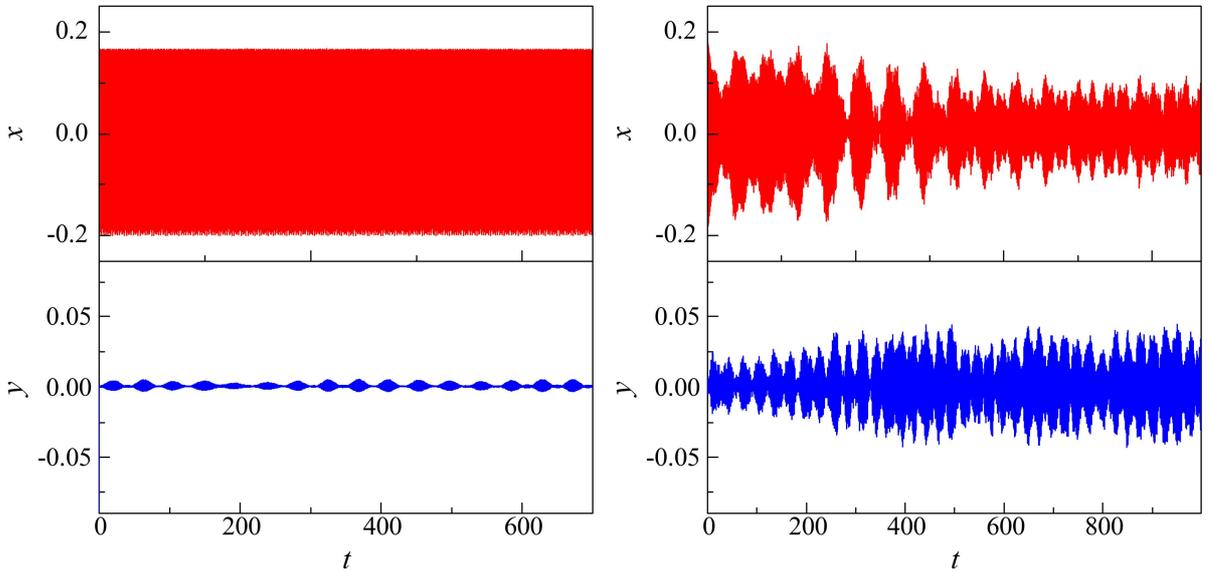

Fig. 2. Relative position $x$ and $y$ (in Å; $x$ and $y$ axes are chosen along the armchair and zigzag directions, respectively) of the centers of mass of the graphene layers as a function of time $t$ (in ps) calculated using the developed potential at temperature of 4.2 K: (a) the two infinite layers, (b) the graphene flake on the infinite layer.

The small Q-factor values of the graphene-based nanoresonator are related to the intense energy exchange of the considered vibrations of the graphene layers with other vibrational modes of the system. It is seen from Fig. 2 that the translational vibration of the layers in the perpendicular direction is easily excited. This is due to the degeneracy of the vibrations in the perpendicular directions, which is a result of the graphene symmetry. The excitation of the vibrations in the perpendicular direction is an intrinsic property of the graphene system and is not sensitive to the choice of the potential.

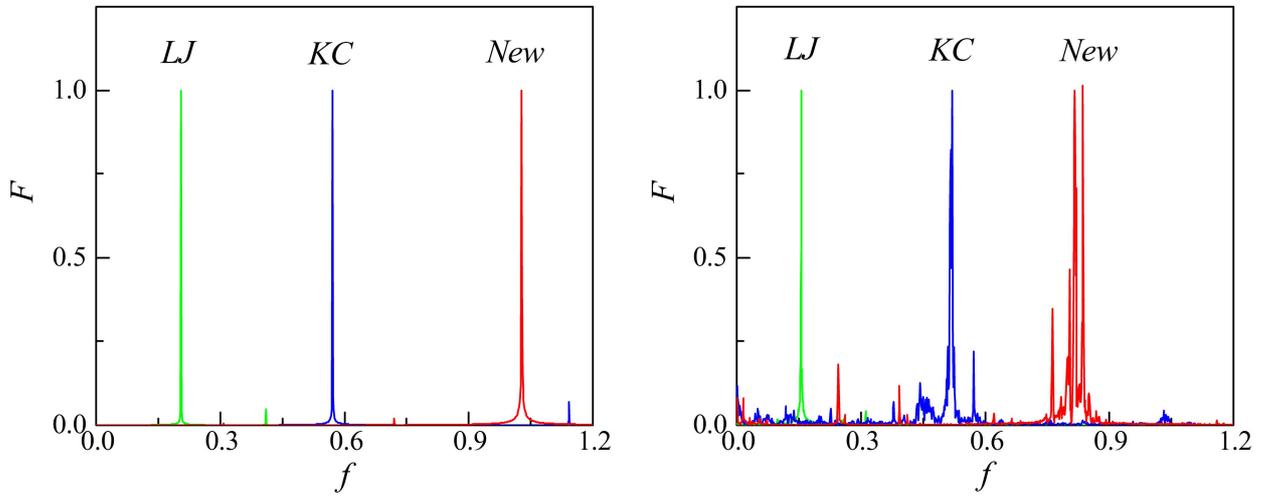

Fig. 3. Calculated Fourier transforms of the relative displacement of the graphene layers along the armchair direction at temperature 4.2 K: (a) the two infinite layers, (b) the graphene flake on the infinite layer. LJ (green line): Lennard-Jones potential, KC (blue line): Kolmogorov-Crespi potential, New (red line): developed potential. Frequency $f$ is given in THz.

Furthermore, the high dissipation in the graphene-based nanoresonator is provided by the excitation of other low frequency vibrational modes, such as the flexural vibrations of the graphene layers and torsional vibrations of the flake. The fundamental frequency of the flexural vibrations of the layers can be found as

$$f_b = \frac{2\pi c_b}{L^2}, \qquad (3)$$

where $c_b$ was found to be $c_b = 5.6 \cdot 10^{-7}\,\text{m}^2/\text{s}$ (see Ref. [50]) and $L$ is the length of the graphene layer. The effective excitation of the flexural vibrations should be observed at $f_b \leq f$, i.e. for lengths $L > 2\,\text{nm}$ for the developed potential and $L > 3\,\text{nm}$ for the Kolmogorov-Crespi potential, which is the case in our MD simulations.

Table 2. Calculated frequency $f$ and Q-factor $Q$ of the graphene-based nanoresonator at temperature 4.2 K.

| Potential | New | Kolmogorov-Crespi | Lennard-Jones |
|---|---|---|---|
| | two infinite layers | | |
| $f$ (THz) | $1.0278 \pm 0.0005$ | $0.5811 \pm 0.0004$ | $0.2051 \pm 0.0004$ |
| $Q$ | $150 \pm 80$ | $110 \pm 80$ | $40 \pm 30$ |
| | graphene flake on infinite graphene layer | | |
| $f$ (THz) | $0.817 \pm 0.002$ | $0.520 \pm 0.002$ | $0.1562 \pm 0.0003$ |
| $Q$ | $33 \pm 20$ | $20 \pm 3$ | $43 \pm 20$ |

It should be also mentioned that for the graphene flake on the substrate graphene layer, the frequency of the in-plane rotational vibrations of the flake about the center of mass is close to the frequency of translational vibrations. This should be valid for any square flake, while for elongated flakes, the resonance between the rotational and translational vibrations should be avoided. The excitation of the rotational vibrations is clearly observed in the MD simulations with the Lennard-Jones potential. In the simulations with the developed potential and Kolmogorov-Crespi potential, the structures of the graphene flake and the substrate graphene layer show a noticeable roughness, which prevents rotation of the flake.

The degeneracy of the translational vibrations in the perpendicular directions, fast energy transfer to the rotational and flexural vibrations of the graphene layers provide relatively small

Q-factor values for the small translational vibrations of the layers. This is opposed to carbon nanotubes for which high Q-factor values up to 100-1000 were reported (see Ref. [16]). Carbon nanotubes are one-dimensional structures, so the translational vibration of the movable wall along the axis is not degenerate. For a (5,5)@(10,10) nanotube, the frequencies of the translational vibrations of the movable wall along the axis and of the rotational vibrations about the axis can be close [51]. However, this resonance is avoided for all other nanotubes [51]. Since nanotubes are stiffer than graphene, the flexural vibrations of nanotubes become important for the energy dissipation only at long nanotube lengths [48]. This explains high Q-factor values for the nanotube-based nanoresontators as compared to the graphene-based one.

In addition to the graphene-based nanoresonator, we examined the possibility to design the gigahertz oscillator based on the telescopic oscillation of graphene layers [3]. In these simulations, a stack of 5 graphene flakes of 3 nm x 3 nm size was considered. The upper and lower graphene flakes were fixed. The middle graphene flake was pulled out at 3 nm and released with the zero center-of-mass velocity. The layer was observed to retract back into the stack. However, no oscillations of the movable flake occurred. This means that the Q-factor of the graphene-based gigahertz oscillator is very low ($Q<1$). We believe that this result is related to the fact that the magnitude of corrugation of the interlayer interaction energy against sliding for commensurate graphene layers is of the order of the interlayer interaction energy itself.

## 4. Discussion

Let us discuss the possibility to design NEMS based on the relative motion or telescopic oscillation of graphene layers. Electromechanical nanorelays based on the telescopic extension of a shell of inner walls of multi-walled carbon nanotubes were fabricated recently [21, 22]. By analogy, we propose nanorelays based on bilayer or multilayer graphene. Similar to nanorelays based on carbon nanotubes (see Ref. [52] for a review), graphene-based nanorelays can operate with and without the gate electrode and can be used as memory cells.

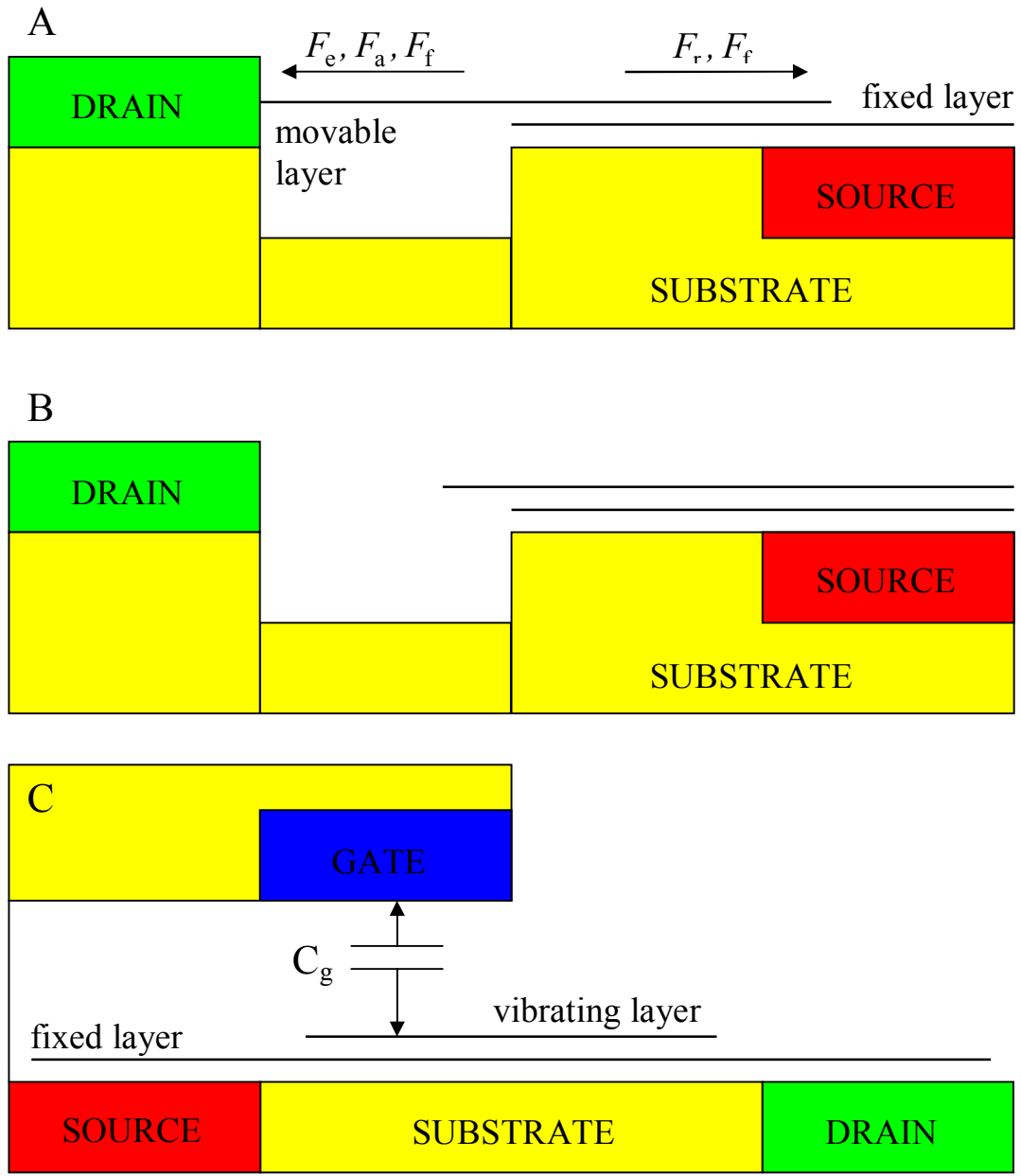

Fig. 4. Schemes of graphene-based NEMS. A: nanorelay (position 'on'), B: nanorelay (position 'off'), C: nanoresonator based on the relative vibrations of graphene layers.

The operation of the graphene-based nanorelay shown in Fig. 4A,B is determined by the balance of forces applied to the movable layer (or several movable layers) of bilayer or multilayer graphene attached to the source electrode. These forces are the attraction force $F_a$ of the van der Waals interaction between the movable layer and the drain electrode, the electrostatic force $F_e$ between the movable layer and the drain electrode, as well as the forces of interaction between the movable and fixed layers of bilayer or multilayer graphene attached to the source electrode. The latter comprise the force $F_r$ which acts at the edge of the fixed layer and retracts the movable layer back onto the fixed layer during its extension, and the static friction force $F_f$ which arises from the interaction of the overlapping layers. The position 'off' of the nanorelay (which corresponds to the logical state 0 of the memory cell) is established when the movable layer of the bilayer or multilayer graphene is retracted onto the fixed layer (Fig. 4B). The position 'on' of the nanorelay (which corresponds to the logical state 1 of the memory cell) is the position at which the movable layer is attracted to the drain electrode by the van der Waals or/and the electrostatic forces (Fig. 4A).

Let us discuss the advantages of the proposed graphene-based memory cell in comparison with the analogous memory cell based on telescoping nanotubes. The operating frequency of the memory cell based on the telescopic extension of nanotube walls is restricted by the mechanical oscillations which occur after the switching to the state 1. These oscillations increase the switching time up to 100 ps for memory cells based on telescoping nanotubes, whereas the switching time to the state 0 was theoretically estimated as being less than 1 ps [53]. Note that for the oscillators based on the relative oscillations of walls of a double-walled nanotube, the Q-factor lies in the range of 100 – 1000 (see Ref. [23, 48] and references therein). As opposed to carbon nanotubes, for telescoping graphene layers, our MD simulations showed that no oscillations appear upon retraction of the movable graphene layer. This means that the graphene-based memory cells should be characterized with the smallest possible switching times.

Moreover, for the nanotube-based memory cell, the forces $F_a$ and $F_r$ are determined by the radius of the movable wall and can not be chosen independently. For the graphene-based memory cell, the force $F_a$ is determined by the shape of the edge of the movable layer interacting with the drain electrode and the force $F_r$ is determined by the overlap length of the layers at the edge of the fixed layer. Thus, the forces $F_a$ and $F_r$ can be chosen independently by the design of the shape of the movable layer.

An ultra-high frequency nanoresonator based on the small relative vibrations of nanotube walls was proposed to be used as a mass nanosensor [16]. Here we consider the possibility of the analogous nanoresonator based on the small relative vibrations of graphene layers (Fig. 4C). Our calculations showed that the Q-factor of the graphene-based nanoresonator is $Q < 200$, which is considerably less than the Q-factor of the nanoresonators [16] based on the small relative vibrations of nanotube walls. Therefore, the nanoresonator based on the small relative vibrations of graphene layers has less mass sensitivity than the nanotube-based nanoresonator. Nevertheless, this NEMS can be used to determine the frequency of the small relative vibrations of graphene layers and thus to verify the calculations of the potential relief of interlayer interaction energy.

## 5. Conclusions

*Ab initio* and empirical calculations of the potential relief of interlayer interaction energy in bilayer graphene were performed. A new potential based on the density functional theory calculations with the dispersion correction was developed to reproduce the potential relief of interlayer interaction energy in bilayer graphene. The MD simulations of the telescopic oscillations and small relative vibrations of graphene layers were carried out. It was shown that both telescopic oscillations and small relative vibrations have small Q-factor values. For this reason, graphene is worse for the use in the gigahertz oscillator and nanoresonator as compared to carbon nanotubes. However, this property makes graphene perfect for the use in fast-responding nanorelays and memory cells. Due to the small Q-factor values, no oscillations occur upon switching the position of the movable layer in graphene-based nanorelays and memory cells, providing the smallest possible switching times. It was also suggested that the balance of forces determining the operation of the graphene-based nanorelay can be controlled via the shape of the movable layer. We showed that the nanoresonator based on the small relative vibrations of graphene layers should have smaller mass sensitivity than that of the similar nanotube-based nanoresonator. Nevertheless, the measurements of frequency of such a graphene-based nanoresonator would allow to verify the calculations of the potential relief of interlayer interaction energy in bilayer graphene.

## Acknowledgement

This work has been supported by the RFBR grants 08-02-00685 and 10-02-90021-Bel.